\documentclass[prl, twocolumn,  superscriptaddress]{revtex4-1}

\usepackage{graphicx,subfigure}
\usepackage{dcolumn}
\usepackage{epsfig}
\usepackage{latexsym}
\usepackage{amsfonts}
\usepackage{amssymb}

\begin{document}

\title{Fast Diffusion of Long Guest Rods in a Lamellar Phase of Short Host Particles
	}

\author{Laura Alvarez$^{1,2}$, M. Paul Lettinga$^{2,3,\ddag}$ and Eric Grelet}
\altaffiliation{grelet@crpp-bordeaux.cnrs.fr, 
	$^\ddag$~p.lettinga@fz-juelich.de}
\affiliation{
Centre de Recherche Paul-Pascal, CNRS \& Universit\'e de Bordeaux, 115 Avenue Schweitzer, F-33600 Pessac, France \\
$^2$ Laboratory for Soft Matter and Biophysics, KU Leuven, Celestijnenlaan
200D, B-3001 Leuven, Belgium \\
$^3$ ICS-3, Forschungszentrum
J\"{u}lich, D-52425 J\"{u}lich, Germany \\}

\date{\today}
\begin{abstract}
We investigate the dynamic behavior of long guest rod-like particles immersed in liquid crystalline phases formed by shorter host rods, tracking both guest and host particles by fluorescence microscopy. Counter-intuitively, we evidence that long rods diffuse faster than short rods forming the one-dimensional ordered smectic-A phase. This results from the larger and non-commensurate size of the guest particles as compared to the wavelength of the energy landscape set by the lamellar stack of liquid slabs. 
The long guest particles 
are also shown to be still mobile in the crystalline smectic-B phase, as they generate their own voids in the adjacent layers.
\end{abstract}


\maketitle

A crowded environment usually causes a slowing down of the dynamics of the constituting particles \cite{Banks2005,Dix2008,Sokolov2012}. In the macroscopic world, traffic jams represent a good illustration of frustrated motion due to crowding, while at the microscopic scales frustrated motion is observed for instance in polymeric \cite{DOIEdwards,Hermans1982} and colloidal glasses \cite{Weeks2002,Chaudhuri2007}. 
Generally, the  degree of frustration strongly depends on the size ratio between the frustrated guest particle and the host particles that cause the frustration. When this ratio is very high then the crowding is experienced by an effective viscosity 
and the Stokes-Einstein relation is obeyed \cite{Koenderink2003}. In the opposite case where the guest particles are much smaller, e.g. beads in a filament network, the mesh size of that network sets the degree of frustration \cite{Wong2004,Kang2005,vanderGucht2003}. As a general rule reported in literature, large guest particles are slower than small host particles and small guest particles are faster than large host particles. 
In this paper, we show that this rule does not always hold as large guest particles can be more mobile than the small host particles, depending on the self-organization 
of the latter.

In most studies on crowding, the motion of tracers is tracked in amorphous media \cite{Wong2004,Kang2005,vanderGucht2003,Koenderink2003,Berthier2011,Mukhija2007,Sonn-Segev2014}. Effect of crowding has also been shown in highly ordered phases, in which the single particle dynamics is strongly affected \cite{Kuijk2012,Naderi2013}. For instance, in three dimensional colloidal crystals, dynamics is determined by the local mobility at the crystal lattice points and by the existence of vacancies \cite{Holmqvist2014,Alsayed2005}. Reducing the dimensionality of the positional ordering from 3D to 1D leads to an unexpected hopping-type diffusion \cite{Lettinga2007,Grelet2008a,Patti2010,Naderi2014}. This has been first observed in the model system of filamentous viruses \cite{Dogic2006,Grelet2008} organized in a smectic-A phase, which can be considered as a one-dimensional stack of liquid slabs \cite{Dogic2001,Grelet2014}. Quasi-quantized steps matching the smectic layer spacing $L_{layer}$, have been evidenced using single particle tracking by fluorescent microscopy \cite{Pouget2011}. Here the particles essentially behave as Brownian particles in a one-dimensional potential \cite{Lettinga2007,Pouget2011} as the jumping rates depend on the ordering potential associated with the lamellar organization. 
In biological membranes, which can be seen as single smectic slabs \cite{Brown1979,Wolken1966}, different macromolecules such as proteins, are transported through the membranes, usually formed by lipids \cite{Kusumi1993}. This leads to the question of \textit{how transport of guest particles, which are not part of the structure forming materials, occurs through highly ordered states}.
In most studies where the diffusion of guest particles through a structured background is investigated, the guest particles are smaller than the typical length scale that characterizes the host system, both in simulations \cite{Zwanzig1988,Lindner2013} and in experiments \cite{Blickle2007,Evstigneev2008,Volpe2014,Tierno2016,Evers2013}. Here, we address the opposite limit from an experimental point of view: how is the mobility of a guest particle affected by a surrounding energy landscape that has a smaller length scale than the guest particle?

\begin{figure}
	\begin{center}
		\includegraphics[width=\columnwidth]{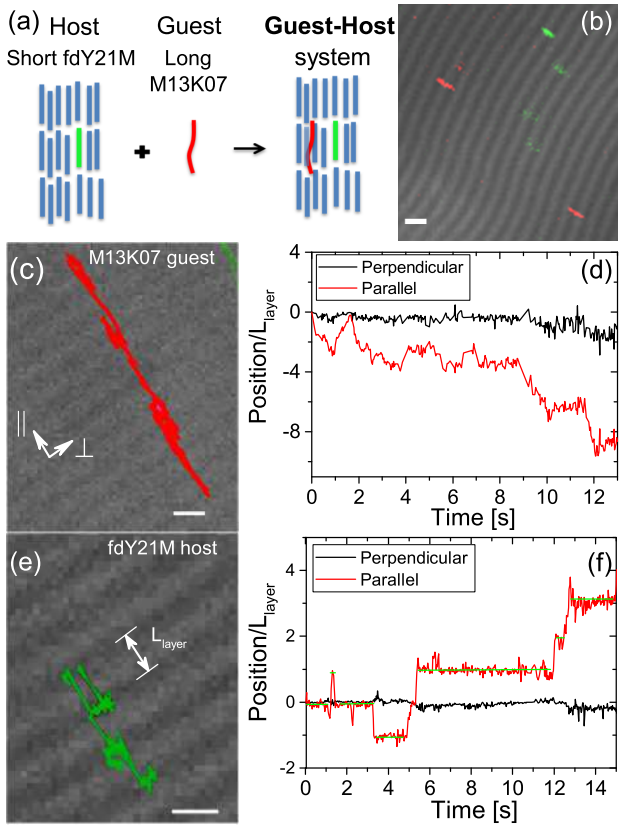}
	\end{center}
	\caption{(a) Schematic representation of the {\it guest-host} system exhibiting a smectic organization: the host is formed by short fdY21M viruses with both tracer amounts of labeled host particles (green, fdY21M) and of long labeled semi-flexible guest rods (red, M13K07), which are 1.3 times longer than the host ones. (b) Overlay of a differential interference contrast (DIC) microscopy image evidencing smectic layers, and of a fluorescence image, displaying the dual labeling of the host and guest particles. (c) Overlay of a DIC picture and a M13K07 guest trajectory for which rapid diffusion is observed through the layers. (d) Corresponding displacements of the M13K07 rod, parallel (red) and perpendicular (black) to the normal of the smectic layer. (e) Example of trace for a fdY21M host particle in the lamellar phase of layer spacing $L_{layer}$. (f) Associated displacement where jumping-type diffusion process is clearly evidenced. The green lines are obtained by the jump recognition algorithm and define the residence time, $\tau_{res}$, that rods spend within a smectic layer before a hopping-type event. Scale bars represent 1 ~$\mu m$.}
	\label{fig_guesthost}
\end{figure}

For this study, we have devised a model system by introducing a tracer amount of long, non-commensurate, guest rods in a host smectic phase comprised from shorter host rods, 
as schematically represented in Fig. ~\ref{fig_guesthost}(a).
Having guest particles which are longer than the host layer spacing for which $L_{layer}\cong L_{host}$ \cite{Grelet2014}, 
implies that the guest rods have to be \textit{accommodated} in more than one smectic layer, 
exceeding the typical length scale of the host ordering potential. 
Filamentous viruses are ideally suited for such fundamental studies, as they are intrinsically monodisperse but can be produced in different lengths \cite{Dogic2001}. Furthermore, their colloidal size allows for their imaging and tracking at the single particle level. 
These charged rod-like particles have been shown to behave nearly as hard rods, by defining an effective diameter accounting for their electrostatic repulsion at different ionic strengths \cite{Grelet2014}.
We will show that the diffusion of the non-commensurate particles through the one-dimensional ordered smectic-A (SmA) and through the crystalline smectic-B (SmB) phase formed by shorter host rods is less sensitive to the smectic potential and that they permeate the layers faster \textit{because} they are longer.

We used two mutants of the filamentous bacteriophage fd to create our \textit{guest-host} model system with fdY21M virus as a short stiff host (contour length  $L_{fdY21M} = 0.91~\mu m$, persistence length  $P_{fdY21M} = 9.9~\mu m$, diameter  $d = 7$~nm) and  M13K07 helper phage as long guest semi-flexible rod ($L_{M13K07} = 1.2~\mu m$,  $P_{M13K07} = 2.8~\mu m$, $d = 7$~nm) \cite{Barry09,Pouget2011,Sharma2014}, both prepared following standard biological protocols \cite{Maniatis}. Consequently, the guest-host length ratio is $L_{M13K07}/L_{fdY21M}\cong L_{M13K07}/L_{layer}= 1.3$, as shown in Fig. ~\ref{fig_guesthost}(b). The choice of fdY21M as stiff rods for the host matrix enables to significantly increase the smectic-A range compared to semi-flexible fd-wt suspensions \cite{Grelet2014}, allowing therefore for a more accurate study of the dynamics close to both the chiral nematic-SmA and SmA-SmB phase transitions.
A second batch of fdY21M viruses has been prepared, for which an extended amplification time of the bacteria cultures has been applied, yielding to the formation of a low fraction of multimeric viruses in the suspension. The virus contour length being fixed by the size of its DNA, a few percents of dimers and trimers (containing respectively two and three times the virus genome) exist in this polydisperse batch (See the Supplemental Material for details \cite{SM}). 
We labeled fdY21M and M13K07 with green and red fluorescent dyes, respectively, to easily distinguish them by fluorescence microscopy. 
Labeled particles were added in a ratio of one labeled particle over $10^5$ non-labeled particles such that trajectories of individual rods can be recorded (Fig. \ref{fig_guesthost}). The amount of labeled particles is small enough to assume that the phase behavior of the host system remains unaffected. The multimeric batch containing fdY21M dimers and trimers in small amount exhibit similar quantitative phase behavior as the batch of pure single host particles (See below). A broad range of concentrations of aqueous suspensions of virus (in TRIS-HCl-NaCl buffer, pH 8.2, ionic strength of 20~mM) has been investigated from the (chiral) nematic to the smectic-A and smectic-B phases \cite{Grelet2014}.  

Figures \ref{fig_guesthost}(c)-(f) show two typical trajectories recorded in the smectic-A phase 
for both M13K07 guest and fdY21M host particles. Qualitatively, two main features distinguish their respective dynamic behavior: (1) The long guests do not exhibit hopping-type events as sharp as those observed for the host particles, which also stay longer in a given smectic layer having therefore a longer residence time, $\tau_{res}$;
(2) The parallel displacement (along the director, or, equivalently along the normal of the smectic layers indicated by the \textbf{z} axis) for a given observation time is bigger for the long M13K07 guest viruses.
We quantify these observations by comparing the ordering potentials that are effectively felt by both tracer particles and by measuring the associated dynamic behavior of the guest and host particles through their mean square displacements (MSD).

\begin{figure} [!ht]
	\begin{center}
		\includegraphics[width=0.9\columnwidth]{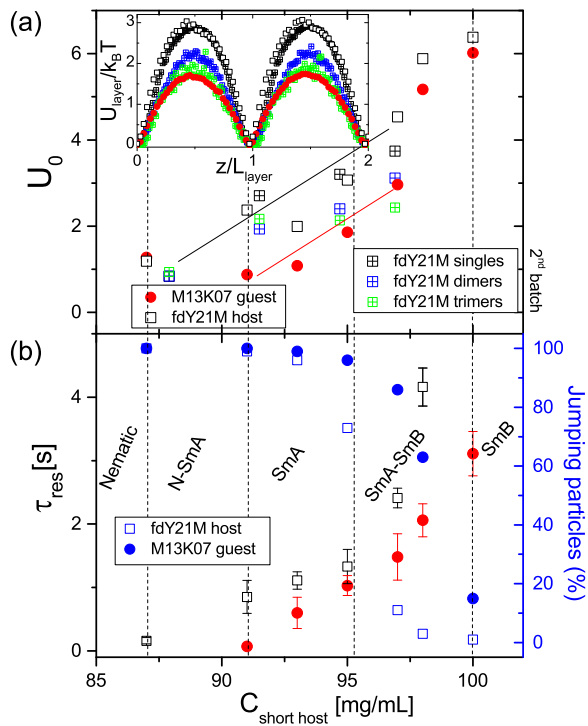}
	\end{center}
	\caption{Concentration dependence of (a) smectic potential barriers obtained after deconvolution from the point-spread function of the microscope 
		for long non-commensurate guest (red solid circles), long commensurate guest (blue and green crossed squares) and host (black open squares and crossed squares) particles. Solid lines are guide to the eye. Inset: Raw data (before deconvolution) of the energy landscape at a host concentration of $C_{fdY21M}=95$~mg/mL. (b) Average residence time $\tau_{res}$ for which the particles remain in a layer before a hopping-type event, and percentage of jumping particles 
		during the acquisition time window of 15~s (blue symbols). The vertical dashed lines indicate the phase transitions. 
	}
	\label{fig_Potentials}
\end{figure}

The 1D ordering potential, $U_{layer}$, which characterizes the lamellar organization of the smectic phase, can be obtained experimentally from the particle fluctuations with respect to the middle of the layer \cite{Lettinga2007}. The distribution of these fluctuations is proportional to the probability $P(z)$ of finding the center-of-mass of the rod-shaped particles with respect to the middle of the layer, via the Boltzmann factor, $P(z)\sim e^{-U_{layer}(z)/k_bT}$ \cite{Pouget2011}.
Fig. \ref{fig_Potentials}(a) displays 
the concentration dependence of the potential barrier $U_0$ for the different particles. 
While for the short host particles the onset of $U_{layer}^{fdY21M}$ occurs at the N-SmA transition, as expected, the onset of $U_{layer}^{M13K07}$ for the long guests is shifted to much higher concentrations. This means that M13K07 guest particles do not feel the underlying periodic potential as strongly as the fdY21M host ones. Despite their commensurability with the layer spacing, multimeric fdY21M viruses (dimers and trimers) exhibit lower potentials than singles, but still higher than M13K07 particles. This can be understood by considering that dimer center-of-mass experiences a maximum of the potential, contrary to singles which are trapped in a minimum. Such an explanation can be extended to trimers, whose particle body feels two maxima of the smectic potential, reducing accordingly the resulting potential felt by these multimeric viruses.  
The difference in potential between guest and host particles persists at the SmA-SmB transition where $U_{layer}^{M13K07}$ continues to increase, while for the short hosts the diffusion freezes in the crystalline smectic-B phase. This corresponds to a divergence of the ordering potential, which impedes its accurate determination. It is worth mentioning that both particles have the same concentration dependence, since the slope is similar for both curves. This is directly reflected in the ratio of particles that jump layers over particles that do not jump  (within the time window of a trajectory) (Fig. \ref{fig_Potentials}(b) in blue):  
hopping-type events can still be found in the SmB phase for the long M13K07 guests at concentrations where the host particles seized to jump. Similarly, the average residence time $\tau_{res}$, which is calculated from the histogram of residence times for all particles, strongly increases around the SmA-SmB transition for the host particles, while $\tau_{res}$ rises more smoothly throughout that transition for the long guest ones, as shown in Fig. \ref{fig_Potentials}(b). Note that for the longest residence time $\tau_{res} \simeq 4~$s, only 3\% of the particles continue to display jumps within the observation time window of 15~s. This confirms that our acquisition time window is long enough to capture, on average, any hopping-type events occurring in the system. 

\begin{figure}
	\begin{center}
		\includegraphics[width=0.9\columnwidth]{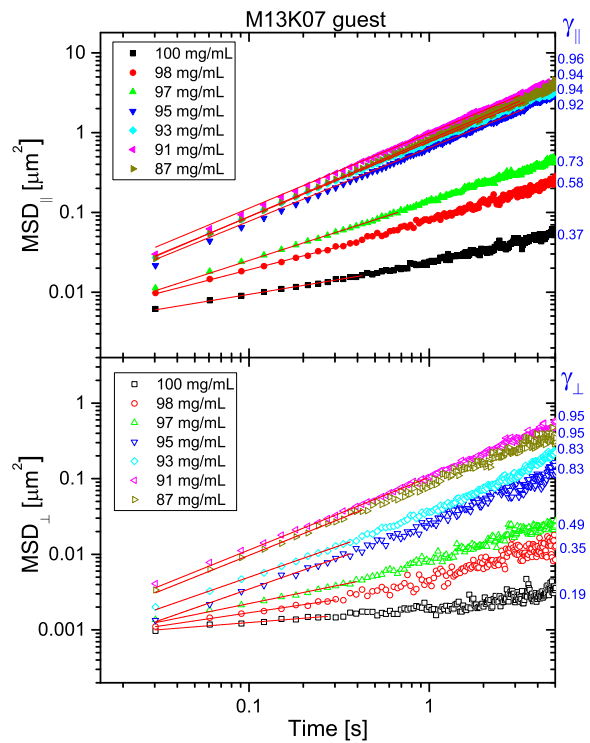}
	\end{center}		
		\caption {Log-log representation of the mean square displacement (MSD) parallel (top) and perpendicular (bottom) to the director plotted as a function of time for the long guest M13K07 particles in the range of high host concentrations. The diffusion exponents $\gamma$ obtained from the numerical fit (red lines) according to Eq. \ref{eq_MSDrods} are also indicated.}
		\label{Fig3}
\end{figure}

\begin{figure} 
	\begin{center}
		\includegraphics[width=0.9\columnwidth]{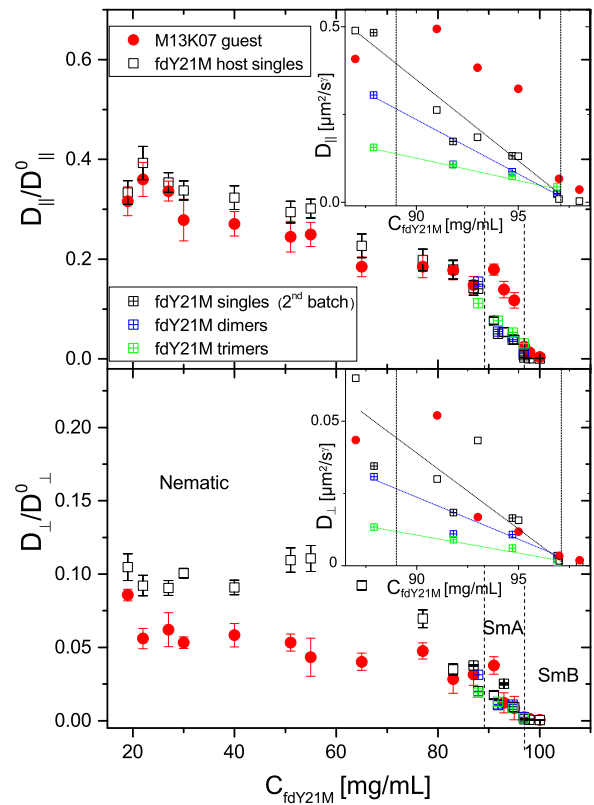}
	\end{center}
		\caption{Diffusion rates \textit{normalized} by the anisotropic diffusion coefficients at infinite dilution $D^0_{\|}$ and $D^0_{\bot}$ (See the Supplemental Material \cite{SM}) 
		for long non-commensurate guest (red solid circles), long commensurate guest (blue and green crossed squares) and host (black open squares and crossed squares) particles. Insets: Zoom-in of the \textit{raw} diffusion rates in the smectic range evidencing that long (dimers and trimers) \textit{commensurate} rods diffuse slower than single host ones, while long \textit{non-commensurate} ones (M13K07) diffuse faster. Dotted lines are a guide for the eye. 
		The vertical dashed lines indicate the phase boundary concentrations. }
		\label{fig_LayerDiffcoeff}
\end{figure}

Beyond the ordering potentials, the self-diffusion of the system can be quantified 
through the full range of concentrations by the mean square displacement (MSD), which is determined from the particle trajectories \textbf{r}(t). In complex systems, the time evolution of the MSD is not always linear as for diffusive particles, but it is more generally described by a power law:

\begin{eqnarray}
	MSD_{\|,\bot}=\langle r_{\|,\bot}^2(t)\rangle=2\times D_{\|,\bot}t^{\gamma_{\|,\bot}},
	\label{eq_MSDrods}
\end{eqnarray}

\noindent where $D_{\|,\bot}$ and $\gamma_{\|,\bot}$ are the diffusion rates and exponents parallel ($\|$) and perpendicular (${\bot}$) to the director, respectively. 
The anisotropic MSDs of the guest particles are displayed in Fig. \ref{Fig3} for the highest concentrations corresponding mainly to the smectic organization.
Diffusion rates and exponents have been obtained from a fit of these data according to Eq. \ref{eq_MSDrods}.
Diffusive behavior is obtained for both rod systems and in both directions in the entire nematic range. 
We do observe that the diffusion rates of the long guest is lower compared to the short host, especially for $D_\bot^{M13K07}$ in the perpendicular direction, as shown in Fig \ref{fig_LayerDiffcoeff}, even when renormalizing by the anisotropic diffusion coefficients at infinite dilution which partially accounts for some rod size effects (See the Supplemental Material \cite{SM}). Thus, in the nematic phase, long rods are more hindered, as one would have expected.
 Inversely, after the N-Sm transition, $D_{\bot}$ is similar for both rods, while
$D_{\|}^{fdY21M}$ for the short hosts sharply decreases (Fig \ref{fig_LayerDiffcoeff}). For fdY21M singles, dimers and trimers, the scaling by their length of the diffusion rates works remarkably well in the smectic-A range. This confirms that the longer the \textit{commensurate} particles are, the slower their diffusion is, as also directly demonstrated in the insets of Fig. \ref{fig_LayerDiffcoeff}.  
This is in strong contrast to the non-commensurate M13K07 guests, for which $D_{\|}^{M13K07}$ even displays a slight increase after the phase transition, showing that \textit{the long  non-commensurate guest particles diffuse significantly faster in the smectic-A phase}. 
The behavior is sub-diffusive, whatever the direction, when approaching the SmB phase. It is important to stress that in this high concentration range, 
we observe after some time a change in the slope of the log-log MSD for all particles (Fig. \ref{Fig3} and Supplemental Material \cite{SM}). The increase of the slopes, and thus of $\gamma$, is interpreted as a consequence of a void creation in the adjacent smectic layer. This results in the release of the layer constraint of the initially confined particles, 
allowing them to jump. The diffusion exponents $\gamma_{\|}$ are higher for the guest particles, which is interpreted as voids that are more easily generated, as compared to voids that have to be created for the host particles. This phenomenon is more apparent in the dense crystalline smectic-B phase. 

The fact that no significant difference is observed between semi-flexible long guests and stiff short hosts in the perpendicular diffusion is rather unexpected. 
Indeed, it has been shown that self-diffusion within the layers is far more pronounced for stiff rods as compared to more flexible ones of the same size \cite{Pouget2011}. 
The reason is that parallel and perpendicular diffusion are correlated, and that M13K07 guests feel a weaker ordering potential than the constituting host particles. This accounts then for their relative promoted lateral diffusion. As a result, also, the diffusion exponents exhibit a less enhanced sub-diffusivity than the host particles at the same concentration (Fig. \ref{Fig3} and Supplemental Material \cite{SM}). For multimeric fdY21M particles, hindrance in the lateral diffusion increases with rod length (Fig. \ref{fig_LayerDiffcoeff}(b)), as these commensurate rods belong simultaneously to many smectic layers. 


In conclusion, the general assumption that large particles always diffuse slower than small ones is not valid as soon as the length-scale associated with the energy landscape formed by the host particles is smaller than the guest particle size. We proved this effect by evidencing a promoted permeation of non-commensurate long guest rods through self-assembled smectic layers of shorter host particles. This phenomenon can be understood, realizing that the ordering potential barrier in the smectic phase is not a static value but rather the result of rod density fluctuations within the layer. A particle can therefore jump when a transient void exists in the adjacent layer. At higher concentrations, the number of voids decreases and hence the potential barrier increases. As non-commensurate long rods are always present in at least two layers at the same time, they generate their own voids facilitating their permeation. Our work should therefore stimulate further theoretical investigation 
for the generalization of the concept evidenced here and for the design of \textit{fast diffusers in crowded environment} with potential application in cell biology and drug delivery.
  
\begin{acknowledgments}
We acknowledge IdEx Bordeaux (France) for financial support, and we thank Olivera Korculanin and Alexis de la Cotte for their help in using analysis softwares and in preparing the experiments. 
\end{acknowledgments}

\clearpage
	

\begin{references}
	
	\bibitem{Banks2005} D. S. Banks and C. Fradin, Biophys. J. \textbf{89}, 2960 (2005).
	\bibitem{Dix2008} J. A. Dix and A. S. Verkman, Annu. Rev. Biophys. \textbf{37}, 247 (2008).
	\bibitem{Sokolov2012} I. M. Sokolov, Soft Matter \textbf{8}, 9043 (2012).
	\bibitem{DOIEdwards} M. Doi and S. F. Edwards, The Theory of Polymer Dynamics (Oxford University Press, USA, 1986).
	\bibitem{Hermans1982} J. Hermans, J. Chem. Phys. \textbf{77}, 2193	(1982).
	\bibitem{Weeks2002} E. R. Weeks and D. A. Weitz, Chem. Phys. \textbf{284}, 361 (2002).
	\bibitem{Chaudhuri2007} P. Chaudhuri, L. Berthier, and W. Kob, Phys. Rev. Lett. \textbf{99}, 060604 (2007).
	\bibitem{Koenderink2003} G. H. Koenderink, H. Y. Zhang, D. G. A. L. Aarts, M. P. Lettinga, A. P. Philipse, G. N\"agele, Faraday Discuss. \textbf{123}, 335 (2003).  
	\bibitem{Wong2004} I. Y. Wong, M. L. Gardel, D. R. Reichman, E. R. Weeks, M. T. Valentine, A. R. Bausch, D. A. Weitz, Phys. Rev. Lett., \textbf{92} 178101 (2004).  
	\bibitem{Kang2005} K. Kang, J. Gapinski, M. P. Lettinga, J. Buitenhuis, G. Meier, M. Ratajczyk, J. K. G. Dhont, A. Patkowski, J. Chem. Phys. \textbf{122}, 044905 (2005).
	\bibitem{vanderGucht2003} J. van der Gucht, N. A. M. Besseling, W. Knoben, L. Bouteiller, M. A. Cohen Stuart, Phys. Rev. E \textbf{67} 051106 (2003).   
	\bibitem{Berthier2011} L. Berthier, Physics \textbf{4}, 42 (2011).
	\bibitem{Mukhija2007} D. Mukhija, M. J. Solomon, J. Colloid Interface Sci. \textbf{314}, 98 (2007).
	\bibitem{Sonn-Segev2014} A. Sonn-Segev, A. Bernheim-Groswasser and Y. Roichman, Soft Matter \textbf{10}, 8324 (2014).
	\bibitem{Kuijk2012}	A. Kuijk, D. V. Byelov, A. V. Petukhov, A. van Blaaderen, and A. Imhof, 
	Faraday Discuss. \textbf{159}, 181 (2012).
	\bibitem{Naderi2013} S. Naderi, E. Pouget, P. Ballesta, P. van der Schoot,	M. P. Lettinga, and E. Grelet, Phys. Rev. Lett. \textbf{111}, 037801 (2013).
	\bibitem{Holmqvist2014} P. Holmqvist, Langmuir \textbf{30}, 6678 (2014).
	\bibitem{Alsayed2005} A. M. Alsayed, M. F. Islam, J. Zhang, P. J. Collings, A. G. Yodh, Science \textbf{309}, 1207 (2005).
	\bibitem{Lettinga2007} M. P. Lettinga and E. Grelet, Phys. Rev. Lett. \textbf{99}, 197802 (2007).
	\bibitem{Patti2010} A. Patti, D. El Masri, R. van Roij, and M. Dijkstra, J. Chem. Phys. \textbf{132}, 224907 (2010).
	\bibitem{Naderi2014} S. Naderi, and P. van der Schoot, J. Chem. Phys. \textbf{141}, 124901 (2014).
	\bibitem{Grelet2008a} E. Grelet, M. P. Lettinga, M. Bier, R. van Roij, and P. van der Schoot, J. Phys. Cond. Matt. \textbf{20}, 494213 (2008).
	\bibitem{Dogic2006} Z. Dogic and S. Fraden, Curr. Opin. Colloid Interface Sci. \textbf{11}, 47 (2006).
	\bibitem{Grelet2008} E. Grelet, Phys. Rev. Lett. \textbf{100}, 168301 (2008).
	\bibitem{Dogic2001} Z. Dogic and S. Fraden, Philos. Trans. R. Soc. London, Ser. A \textbf{359}, 997 (2001).
	\bibitem{Grelet2014}  E. Grelet, 
	Phys. Rev. X {\bf 4}, 021053 (2014).
	\bibitem{Pouget2011} E. Pouget, E. Grelet, and M. P. Lettinga, Phys. Rev. E \textbf{84}, 041704 (2011).	
	

	\bibitem{Brown1979} G. H. Brown and J. J. Wolken, Liquid Crystals and Biological Structures (Academic Press, New York, 1979).
	\bibitem{Wolken1966} J. J. Wolken, J. Am. Oil Chem. Soc. \textbf{43}, 271 (1966).
	\bibitem{Kusumi1993} A. Kusumi, Y. Sako, and M. Yamamoto, Biophys. J. \textbf{65}, 2021 (1993).
	\bibitem{Zwanzig1988} R. Zwanzig, Proc. Natl. Acad. Sci. USA \textbf{85}, 2029 (1988).
	\bibitem{Lindner2013} M. Lindner, G. Nir, A. Vivante, I. T. Young, and Y. Garini, Phys. Rev. E \textbf{87}, 022716 (2013).
	\bibitem{Blickle2007} V. Blickle, T. Speck, U. Seifert, and C. Bechinger, Phys. Rev. E \textbf{75}, 060101(R) (2007).
	\bibitem{Evstigneev2008} M. Evstigneev, O. Zvyagolskaya, S. Bleil, R. Eichhorn, C. Bechinger, and P. Reimann, Phys. Rev. E \textbf{77}, 041107 (2008).
	\bibitem{Volpe2014}  G. Volpe, G. Volpe, and S. Gigan, Sci. Rep. \textbf{4}, 3936 (2014).
	\bibitem{Tierno2016} P. Tierno and M. R. Shaebani, Soft Matter \textbf{12}, 3398 (2016).
	\bibitem{Evers2013} F. Evers, R. D. L. Hanes, C. Zunke, R. F. Capellmann, J. Bewerunge,	C. Dalle-Ferrier, M. C. Jenkins, I. Ladadwa, A. Heuer, R. Castaneda-Priego,
	and S.U. Egelhaaf, Eur. Phys. J. Special Topics \textbf{222}, 2995 (2013).
	\bibitem{Barry09} E. Barry, D. Beller and Z. Dogic, 
	Soft Matter {\bf 5}, 2563 (2009).
	\bibitem{Sharma2014} P. Sharma, A. Ward, T. Gibaud, M. F. Hagan and Z. Dogic, Nature \textbf{513}, 77 (2014). 
	\bibitem{Maniatis} T. Maniatis, J. Sambrook, and E. Fritsch, Molecular Cloning (Cold Spring Harbor
	Laboratory, 1989).
	\bibitem{SM} See Supplemental Material at http://link.aps.org/ which includes Refs. \cite{ChapterSeth,Croker1996,Tirado1979,Burgers} for the details of the experimental methods, the anisotropic translational diffusion coefficients at infinite dilution, the whole set of MSD for guest M13K07 and fdY21M (single hosts, dimers, and trimers) particles, 
	 and for the movies displaying the motion of long guest and short host rods in the smectic phase.
	\bibitem{ChapterSeth} Z. Dogic and S. Fraden, in Soft Matter, ed. G. Gompper and M. Schick, Wiley-VCH, Weinheim, 2006, vol. 2, pp. 1–86.
	\bibitem{Croker1996} J. C. Crocker and D. G. Grier, J. Colloid Interface Sci. \textbf{179}, 298 (1996). 
	\bibitem{Tirado1979} M. M. Tirado, J. G. de la Torre, 
	J. Chem. Phys. \textbf{71}, 2581 (1979).
	\bibitem{Burgers} J. M. Burgers, \textit{On the Motion of Small Particles of Elongated Form Suspended in a Viscous Liquid}, Amsterdam Academy of Science, Nordeman, Amsterdam, 1938.


%
%
%
%


\end{references}
\end{document}